\documentstyle[12pt]{article}
\begin{document}
\thispagestyle{plain}
\begin{flushright}
UNIL-TP-4/95,PRL-TH-95/6 \\[-2mm]
hep-ph/9505239 \\[-2mm]
\end{flushright}
\begin{center}
{\bf Helicity-Flip Bremsstrahlung and the Measurement of CP-Violating
Form Factors in Polarized $e^+e^-$ Collisions}\\
B. Ananthanarayan\\[-2mm]
{\small Institut de Physique Th\'eorique, Universit\'e de Lausanne\\[-2mm]
CH 1015 Lausanne, Switzerland}\\[-2mm]
Saurabh D. Rindani\\[-2mm]
{\small Theory Group, Physical Research Laboratory\\[-2mm]
Navrangpura, Ahmedabad 380009, India}\\[-2mm]
\end{center}
\begin{abstract}
Certain momentum correlations in the production and subsequent decay of
heavy fermion pairs in  $e^+e^-$ collisions have increased sensitivity
to CP-violating electric and ``weak" dipole
form factors of the fermion in the presence of longitudinal polarization
of the $e^+e^-$ beams. However, unless the polarizations of $e^+$ and $e^-$
are equal and opposite, collinear initial-state radiation accompanied by
helicity flip results in a CP-invariant contribution to these
and acts as a background, and must be calculated and
subtracted in order to isolate the CP-violating contribution
due to the dipole form factor.
We calculate this background contribution
to such correlations, to order $\alpha$,
in $e^+e^-\rightarrow \tau^+\tau^-$ and $e^+e^-\rightarrow
t\bar{t}$.  For the anticitipated
luminosities at
$\tau$-charm factories and SLC,
 for the  $\tau$, and at a future linear
collider with centre-of-mass energy of 500 GeV for the top,
these backgrounds are smaller than the statistical background
and may be safely neglected.
\end{abstract}
\newpage
\section{Introduction}
An interesting avenue that may lead us to new physics is through indirect
methods such as the observation of CP violation outside the $K\overline{K}$
(and $B\overline{B}$) systems~\cite{CPV}.
For instance, electric-dipole or ``weak"-dipole type couplings of quarks and
leptons are predicted in the standard model to be too small to be observable
in current experiments. They could be looked for through CP-odd observables
in the production and decay of these particles. For example, non-zero CP-odd
correlations amongst the momenta of certain final-state particles in the
production of $\tau^+\tau^-$~\cite{Ber,Ana1,Ana2}
(and of $ t\bar{t}$ pairs~\cite{tt,Cuy}),
 in $e^+e^-$ collisions, and their subsequent decay
would imply non-zero values for CP-violating form factors that describe the
coupling of the $\tau^+\tau^-$ (and of the $t\bar{t}$) pair to either or both
the photon and the $Z$. CP-odd correlations have been looked for at the LEP
experiments at CERN, and an upper bound has been placed on the weak dipole
coupling of  $\tau^+\tau^-$ to $Z$~\cite{explimit}.
While predictions exist for a variety of CP-odd correlations for future
experiments~\cite{Ber,tt}, certain ``vector" correlations have been shown
to be substantially enhanced when either or both of the incoming beams are
longitudinally polarized~\cite{Ana1,Ana2,Cuy}. It is also found in the case of
$e^+e^-\rightarrow t\bar{t}$ that CP-violating asymmetries in the angular
and energy distributions of leptons arising from $t$ and $\bar{t}$ decay
are enhanced in the presence of longitudinal polarization, increasing
the overall sensitivity to the measurement of the CP-violating dipole
form factors~\cite{Poulose}.

While the fact that longitudinal beam polarization improves the sensitivity
of CP-odd correlations and asymmetries to the underlying CP-violating couplings
recommends the use of polarization for the measurement of these couplings,
there is in principle, a problem at linear colliders, where only electrons
can be easily polarized. In the case when the $e^+$ and $e^-$ polarizations
are not equal and opposite, the initial state is not CP even, and correlations
which are CP odd in the absence of polarization are not necessarily CP odd
in the presence of polarization. In such a case, there would be a contribution
to the correlation even from the CP-invariant interaction terms. This has to
be calculated and subtracted out before the CP-violating parameters can be
extracted from experiment.

In practice, this is not a problem at the lowest
order in the fine-structure constant $\alpha$. This is because for $e^+e^-$
annihilation into a virtual $\gamma$ or $Z$, the contribution to any final
state in the limit of vanishing electron mass $m_e$
comes from a CP-even initial state consisting of opposite-helicity
combinations of $e^+$ and $e^-$.
Thus the correction to the correlations coming from CP-conserving interactions
is negligibly small. However, to the next order in $\alpha$, there is a
contribution from states with like-helicity $e^+$ and $e^-$ because collinear
emission of a hard photon from $e^+$ or $e^-$ can flip its helicity even
in the limit $m_e\rightarrow 0$~\cite{helflip}. This correction has to be
calculated when $e^+$ and $e^-$ longitudinal polarizations are not equal
and opposite, and subtracted from measured correlations to get the
CP-violating component proportional to dipole form factors.

In earlier work~\cite{Ana1,Cuy} on measurement of $\tau$ and $t$
dipole form factors in the presence of
longitudinal beam polarization, the CP-even correction to correlations
was argued to be small. However, no explicit calculation has been carried
out.
The purpose of this note is to calculate for arbitrary $e^+$ and $e^-$
polarizations, the order-$\alpha$ CP-invariant
contribution,
coming from collinear initial-state bremsstrahlung, to certain correlations
in $e^+e^-\rightarrow \tau^+\tau^-$ and
$e^+e^-\rightarrow t\bar{t}$, whose measurement in experiments with
longitudinally polarized beams was advocated earlier~\cite{Ana1,Ana2,Cuy}.
The calculation has been done using the approach
of Falk and Sehgal~\cite{Fal}, who show that in the limit of vanishing
$m_e$, the helicity-flip bremsstrahlung process may be calculated from
the original process by folding it with a universal function
$D_{hf}=(\alpha/ 2\pi)x$, which
describes the emitted radiation. Here $x$ is the ratio
of the energy of the photon to the beam energy.

The approach we will follow here is to explicitly  compute
the contribution of
the helicity-flip bremsstrahlung to correlations of certain observables of
interest, $O_1$
and $O_2$, which are constructed out of the momenta of certain final-state
particles and are CP odd in the absence of beam polarization.
These are sensitive to the real and imaginary parts of the form
factors since they are CPT even and odd, respectively, again in the
absence of beam polarization. However, since $O_1$ is T odd, its correlation
can
get a contribution from a CP-conserving and T-conserving piece of the
interaction only if the amplitude has an absorptive part. Thus, helicity-flip
bremsstrahlung at the tree level cannot contribute to $\langle O_1\rangle$.
This we find to be borne out by our explicit calculation. The CP-even
background
to $\langle O_2\rangle$ however survives.

Our calculations
and numerical results are presented for experimental situations
envisaged for high-precision studies of the $\tau$ system, for instance at the
$\tau$-charm factories ($\tau$CF); for the current SLC experiment;
and at a future collider (the Next Linear Collider) that
would produce $t\bar{t}$ pairs copiously.

The magnitude of these order-$\alpha$ contributions to correlations will decide
the accuracy to which electric and weak dipole form factors can be measured
from an experimental determination of the correlations. The
correlations determined experimentally, with the order-$\alpha$ CP-even
correction subtracted, have to be at least as large as
the square root of the variance of the observable times $1/\sqrt{N}$, where
$N$ is the number of events in the particular final state in question.
This would correspond to the one-standard-deviation upper limit that given
statistics would permit on the dipole form factors. It will generally be
expected that the order-$\alpha$ corrections we calculate here would be
small compared to the zero-order square-root-variance. Then, if it continues
to be small compared to the square-root-variance times $1/\sqrt{N}$, its effect
can be neglected. This happens to be true for all the cases considered here.

Apart from the fact that the corrections are of order $\alpha$, there are
other reasons for them to be small in certain cases. They are small at the
lower of the
$\tau$CF energies due to the subdominant parity-violating contributions from
$Z$ exchange and at SLC due to the non-resonant nature of the bremsstrahlung
effects. They are significantly larger at the higher of $\tau$CF energies
and in the $t\bar{t}$ situation since
neither of the causes mentioned above hold here.

\section{Notation and Formalism}

In this section we will primarily be concerned with the reactions
\begin{equation}
e^-(p_e)e^+(p_{\bar{e}})\to \gamma^*, Z^*\to
\tau^+\tau^-\to \bar{\nu}_\tau X_{\bar{A}}(q_{\bar{A}})\nu_\tau X_B(q_B),
\end{equation}
and
\begin{equation}
e^+e^-\to \gamma^*, Z^*\to t\bar{t}\to W^+bW^-\bar{b},
\end{equation}
where in the former case $X_A$ and $X_B$ can be one of several final states
arising in $\tau$ decay, and the quantities in parentheses
denote four-momenta of particles.  We concentrate on
$A,B=\pi,\rho$ as in Ref.\cite{Ana1,Ana2},
characterized by their resolving power $\alpha$ for the $\tau$
polarization by $\alpha_\pi=1$ and $\alpha_\rho=0.46$ and
since the calculations can be performed analytically.
The reaction (2) is
characterized by $\alpha_b=(m_t^2-2m_W^2)/(m_t^2+2m_W^2)$ and by the fact
that the detected particle has a
mass negligible compared to the parent in contrast to reaction (1). This leads
to simple kinematic replacements in our final computations. Now, the CP-odd
observables whose correlations are of interest are
\begin{equation}
O_1\equiv {1\over 2}\left({\bf \hat{p}}\cdot({\bf q}_{\bar{B}}\times {\bf
q}_A)+
{\bf \hat{p}}\cdot({\bf q}_{\bar{A}}\times {\bf q}_B)\right)
\end{equation}
and
\begin{equation}
O_2\equiv {1\over 2}\left({\bf\hat{p}}\cdot({\bf q}_A+ {\bf q}_{\bar{B}})+
{\bf \hat{p}}\cdot({\bf q}_{\bar{A}}+ {\bf q}_B)\right),
\end{equation}
where $p=\frac{1}{2}(p_{\bar{e}}-p_e)$, and ${\bf \hat{p}}={\bf p}/\vert {\bf
p}\vert$
for (1). For (2), the same expressions can be used, but with the
simplification that  the momenta $q_A$ and $q_B$ represent the single momentum,
viz.,
that of the b quark. Similarly, the $q_{\bar{A}}$ and $q_{\bar{B}}$ both
correspond to the momentum of $\bar{b}$.
The mean values of these due to possible non-vanishing of CP-violating
form factors and
the variance due to standard-model interactions for arbitrary $e^+$ and $e^-$
polarizations have been computed explicitly
and in closed form earlier~\cite{Ana2,Cuy}.
We will now present a schematic discussion
on performing the computation of interest, viz., calculation of $\langle
O_{1,2}
\rangle$ for arbitrary beam polarizations arising due to helicity-flip
collinear
initial-state bremsstrahlung.

The polarized differential cross section written in terms of the
cross sections for definite helicity combinations of $e^+$ and $e^-$
is given by
\begin{eqnarray}
{\rm d}\sigma^P&=& {1\over 4}\left[{\rm d}\sigma_{LL}(1-P_e)(1-P_{\bar{e}})+
{\rm d}\sigma_{LR}(1-P_e)(1+P_{\bar{e}})\right.\nonumber\\
&&\left.+{\rm d}\sigma_{RL}(1+P_e)(1-P_{\bar{e}})
+{\rm d}\sigma_{RR}(1+P_e)(1+P_{\bar{e}})\right].
\end{eqnarray}
where the first subscript stands for the $e^-$ helicity and the second for
that of the $e^+$. $P_e$ and $P_{\bar{e}}$ denote  the
degree of polarization of $e^-$ and $e^+$ respectively.
Our convention is that the polarization is positive for each particle
when it is right-circularly polarized.

With only standard model couplings and when $m_e/\sqrt{s}\to 0$, the
unpolarized cross section receives only the $LR$ and $RL$ contributions.
However, the inclusion of helicity-flip bremsstrahlung introduces an
$O(\alpha)$ correction to ${\rm d}\sigma^P$ through the non-vanishing of the
differential cross section from the $LL$ and $RR$ contributions, even in the
limit $m_e/\sqrt{s}\to 0$. In particular we have the result for the extra
order-$\alpha$ contribution in the equivalent
particle approach~\cite{Fal}:
\begin{equation}
{\rm d}\hat{\sigma}_{LL}(p_e,p_{\bar{e}})=\int_{\xi_{\rm min}}^1{\rm
d}\xi{\alpha\over 2\pi}(1-\xi)\left[{\rm d}\hat{\sigma}_{RL}(p'_e,p_{\bar{e}})
+{\rm d}\hat{\sigma}_{LR}(p_e,p'_{\bar{e}})
\right],
\end{equation}
for the bremsstrahlung corrected reaction
\[ e_{L(R)}^-e_{L(R)}^+ \to f\bar{f}\gamma, \]
where $(1-\xi)$ is the fraction of the beam momentum carried by the collinear
photon, $\xi_{\rm min}=4m_f^2/s$, and $\hat{\sigma}$ denotes the cross section
for the basic process $e^+e^-\rightarrow f\bar{f}$. The two terms in the
square bracket on the right-hand side correspond to the cases when the
collinear
photon is emitted from the $e^-$ and the $e^+$, respectively,  made
explicit by the prime
on their respective momenta $p_e$ and $p_{\bar{e}}$ to denote a degraded
momentum: $p'_{e(\bar{e})}=\xi p_{e(\bar{e})}$.

The corresponding expression
for the right-handed helicities is
\begin{equation}
{\rm d}\hat{\sigma}_{RR}(p_e,p_{\bar{e}})=\int_{\xi_{\rm min}}^1{\rm
d}\xi{\alpha\over 2\pi}(1-\xi)\left[{\rm d}\hat{\sigma}_{LR}(p'_e,p_{\bar{e}})
+{\rm d}\hat{\sigma}_{RL}(p_e,p'_{\bar{e}})
\right].
\end{equation}
The other two helicity combinations correspond to CP-invariant initial
states, and would give a vanishing expectation value for the CP-odd variables
in the absence of CP-violating interactions, and would therefore
not contribute to the background we wish to calculate.

In the presence of CP violation
the differential cross section receives a contribution
proportional to the CP-violating form factors~\cite{Ber} that yields
a non-vanishing result for the expectation values $\langle O_1\rangle$ and
$\langle O_2\rangle$. The task here is to estimate $\langle O_1\rangle$ and
$\langle O_2\rangle$ in the standard model due to (6) and (7).
This requires the expressions for the spin-density matrix
$\chi_{\rm SM}$ of Ref.~\cite{Ber}
for definite helicity combinations of $e^{\pm}$. We have made the
appropriate changes in the quadratic term involving $V_e^i$ and $A_e^i$ and
the expression for the differential cross section expressed in eq.~(4.3) of
Ref.~\cite{Ber}. We then have the expressions for d$\hat{\sigma}$ with
the helicity combinations as they occur on the right-hand side of (6) and (7).
The correlations in the polarized case can then be related
through these equations to correlations
calculated with $\hat{\sigma}_{LR}$ and
$\hat{\sigma}_{RL}$.

Our strategy for calculating these correlations has been to relate the
observables $O_{1,2}$ by a Lorentz transformation to other observables
in the centre-of-mass frame of $e^+$ and $e^-$ which obtains after one
of them has already radiated a collinear photon. This corresponds to
either ${\bf p'_e}+{\bf p_{\bar{e}}}=0$, or ${\bf p_e}+{\bf p'_{\bar{e}}}=0$,
as the case may be.
In either of these frames, the initial state
corresponding to the subprocess $e^+e^-\to f\bar{f}$ with $LR$ or $RL$
helicity combinations is CP even. Hence only CP-even operators
can have nonzero expectation values in this frame.
It can be seen that in going to this frame from the laboratory frame,
$O_1$ is unchanged, since it contains only momentum
components of the observed momenta transverse to the direction of the
boost. Since $O_1$ is CP odd, it follows that its expectation value
vanishes. It therefore vanishes in the laboratory frame as well.
This is as expected, since $O_1$ is T odd, and cannot have nonzero
expectation value at tree level.

We now discuss the calculation of the order-$\alpha$ component of
$\langle O_2\rangle$. For each term in (4),
we can denote without any confusion the momenta of the final
state charged particles by ${\bf {\bf q_+}}$ and by ${\bf
{\bf q}_-}$ in the $\tau$ system and with the understanding that
these stand for the momenta of the $\bar{b}$ and $b$ in the $t$ system.
We need
to compute $\langle O_2\rangle$
with the simplified definition of $O_2\equiv
{\bf \hat{p}}\cdot({\bf q_+}+{\bf q}_-)$ for our present purposes.
In calculating the expectation value for general polarization,
the contribution of the $LL$ combination of helicities involves,
choosing the $z(3)$ axis along ${\bf \hat{p}}$,
\begin{eqnarray}
{\rm d}\hat{\sigma}_{LL}(p_e,p_{\bar{e}})O_2&=&\int_{\xi_{\rm min}}^1
{\rm d}\xi{\alpha\over 2\pi}(1-\xi)\left[{\rm d}\hat{\sigma}_{RL}
(p'_e,p_{\bar{e}})
\right. \nonumber\\
&&\qquad \qquad \qquad \left.+{\rm d}\hat{\sigma}_{LR}(p_e,p'_{\bar{e}})
\right]({\bf q_+}+{\bf q_-})_3.
\end{eqnarray}
The first term on the right-hand side becomes, on writing in the frame
${\bf p}'_e+{\bf p}_{\bar{e}}=0$,
\begin{eqnarray}
\lefteqn{\left.\int{\rm d}\sigma_{RL}(p'_e,p_{\bar{e}})({\bf q_+}+{\bf q}_-
)_3\right|_{\mbox{\small lab. frame}}} \nonumber\\
\qquad &=& \gamma\left.\int{\rm d}\sigma_{RL}(p'_e,p_{\bar{e}})
\left[({\bf q_+}+{\bf q}_-)_3+\beta(q_+^0+q_-^0)\right]
\right|_{\mbox{\small c. m.}}
\end{eqnarray}
where $\gamma$ and $\beta$ parametrize the Lorentz transformation between the
laboratory frame and the centre-of-mass frame of $e^+e^-$ after
photon emission, and read
\begin{equation}
\beta={1-\xi\over 1+\xi},\qquad \gamma={1+\xi\over 2\sqrt{\xi}}.
\end{equation}
The term on the right-hand side involving the spatial component of
momentum vanishes
because it is CP odd, as argued in the case of $O_1$. Hence, we are only
left with the energy term.
The second term of eq.~(8) is computed with the interchange of the roles of
$e^-$ and $e^+$ with the key difference that the
photon is now emitted in the direction opposite to the one in the former case.
The result is then
\begin{small}
\begin{eqnarray}
& \displaystyle
\left.{\rm d}\sigma_{LL}O_2 =\int_{\xi_{\rm min}}^1\Bigl({\alpha\over 2\pi}
\Bigr)(1-
\xi){1-\xi\over 2\sqrt{\xi}}\left[{\rm d}\sigma_{RL}(\hat{s})+
{\rm d}\sigma_{LR}(\hat{s})\right](q_+^0+q_-^0)\right|_{\mbox{\small c. m.}}. &
\end{eqnarray}
\end{small}
where $\hat{s}=\xi s$.

What remains now is to evaluate the energy correlation in (11) in the
c.m. frame, but with $s$ replaced by $\hat{s}$, and put it together
with corresponding expression for the $RR$ helicity combination. Without
going in to the details of this straightforward but somewhat lengthy
calculation, we state here the result:
\begin{eqnarray}
& \displaystyle
\langle O_2\rangle_{\rm SM}=-{2\over \sigma}(P_e+P_{\bar{e}})
{\alpha\over
2\pi}\int_{\xi_{\rm min}}^1{\rm d}\xi{(1-\xi)^2\over
2\sqrt{\xi}}{4\over 3\hat{x}}\sqrt{1-\hat{x}^2} & \nonumber \\
& \displaystyle
\sum_{i,j}{\hat{s}(V_e^iA_e^j+V_e^jA_e^i)\over (\hat{s}-m_i^2-{\rm i}m_i
\Gamma_i)(\hat{s}-m_j^2-{\rm i}m_j\Gamma_j)}&  \\[-2mm]
& \displaystyle
\left[(E_A^*+E_B^*)\Bigl(V_f^iV_f^j(1+{1\over
2}\hat{x}^2)+A_f^iA_f^j(1-\hat{x}^2)\Bigr)\right. & \nonumber \\
& \displaystyle \left.+{1\over 3}
(\alpha_Aq_A^*+\alpha_Bq_B^*)(A_f^iV_f^j+A_f^jV_f^i)(1-
\hat{x}^2)\right],& \nonumber
\end{eqnarray}
where $\hat{x}=2m_f/\sqrt{\hat{s}}$, $m_i$ and $\Gamma_i$ are respectively
the mass and width of the intermediate vector boson $i$, and $\sigma$ is the
lowest-order cross section up to a normalization factor:
\begin{eqnarray}
& \displaystyle
\sigma ={4\over 3}(1-x^2)^{1\over 2}\sum_{i,j}{s\over
(s-m_i^2+{\rm i}m_i\Gamma_i)(s-m_j^2-{\rm i}m_j\Gamma_j)} & \nonumber\\
& \displaystyle  \left\{(V_e^iV_e^j+A_e^iA_e^j)(1-
P_eP_{\hat{e}})+(V_e^iA_e^j+V_e^jA_e^i)(P_{\bar{e}}-P_e)\right\}&
\nonumber\\
& \displaystyle
\cdot\left\{V_f^iV_f^j\bigl(1+{x^2\over 2}\bigr)+A_f^iA_f^j(1-
x^2)\right\}, &
\end{eqnarray}
with $x=2m_f/\sqrt{s}$. In (12), the asterisks denote the energy and
momenta evaluated in the $\tau$ or $t$ rest frame in the respective cases.
$V_e^i,V_f^i$ are the vector couplings of $e$ and $f$ currents to the boson
$i$, and $A_e^i,A_f^i$ are the corresponding axial vector couplings. The values
of these may be found in \cite{Ber}.

Defining the basic integrals over ${\rm d}\xi$ as
\begin{small}
\begin{eqnarray}
& \displaystyle
J_n^{ij}(s)=\int_{\xi_{\rm min}}^1{\rm d}\xi\ & \nonumber \\
& \displaystyle \cdot{\rm Re}\left[
{(1-\xi)^2\sqrt{\xi-x^2}\over
\bigl(\xi-{m_i^2/ s}+{\rm i}{m_i\Gamma_i/ s}\bigr)
\bigl(\xi-{m_j^2/ s}-{\rm i}{m_j\Gamma_j/
s}\bigr)}\right]\left(\sqrt{\xi}\right)^{1-n},
&
\end{eqnarray}
\end{small}
for $n=0,\ 2,$ we express our final result as
\begin{eqnarray}
& \displaystyle
\langle O_2\rangle_{\rm SM}=-{(P_e+P_{\bar{e}})\over \sigma}\Bigl(
{\alpha\over 2\pi}\bigr){4\over 3xs} & \nonumber\\
& \displaystyle
\sum_{i,j}(V_e^iA_e^j+V_e^jA_e^i)\Bigl[(E_A^*+E_B^*)
\Bigl\{(V_f^iV_f^j+A_f^iA_f^j)J_0^{ij}(s) & \nonumber\\
& \displaystyle
 +\left.\Bigl({1\over 2}V_f^iV_f^j-A_f^iA_f^j\Bigr)
x^2J_2^{ij}(s)\right\}& \nonumber\\
& \displaystyle \left.+{1\over
3}(\alpha_Aq_A^*+\alpha_Bq_B^*)(V_f^iA_f^j+V_f^jA_f^i)
\bigl(J_0^{ij}(s)-x^2J_2^{ij}(s)\bigr)\right]. &
\end{eqnarray}
Note that in the event $P_e=-P_{\bar{e}}$, which can be achieved at circular
colliders to maximize the prospects of detecting CP violation,
$\langle O_2\rangle_{\rm SM}$ is zero. The bremsstrahlung effect is primarily
due to unequal
polarizations and strongly involves parity violating couplings of the gauge
bosons to the electron current.

In Ref.~\cite{Ana1,Ana2} the measurement of the CP-violating form factor by
constructing a polarization asymmetrized distribution
\[ {\rm d}\sigma^A={\rm d}\sigma^P-{\rm d}\sigma^{-P},\]
has been advocated, where $P$ is the effective polarization given by
\[ P=\frac{P_e-P_{\bar{e}}}{1-P_eP_{\bar{e}}}. \]
This is found to enhance the sensitivity of the
measurements. Here we find that $O_2$ receives a standard model contribution
which we denote by $\langle O_2\rangle_{\rm SM}^A$. this is obtained very
simply
from the expression for $\langle O_2\rangle_{\rm SM}$ by rearranging the
expression as $N[P_e,P_{\bar{e}}]/\sigma [P_e,P_{\bar{e}}]$ and obtaining
\begin{equation}
\langle O_2\rangle_{\rm SM}^A = {N[P_e,P_{\bar{e}}]-N[-P_e,-P_{\bar{e}}]\over
\sigma[P_e,P_{\bar{e}}]-\sigma[-P_e,-P_{\bar{e}}]},
\end{equation}
This is independent of $P_e$ and $P_{\bar{e}}$, as the polarization dependence
cancels out between the numerator and denominator.

\section{Numerical Results}

We have carried out numerical evaluations of $\langle O_2\rangle_{\rm SM}$ and
$\langle O_2\rangle_{\rm SM}^A$ for the energies we have considered in
Ref.~\cite{Ana2} for different values of $P_e$ setting $P_{\bar{e}}=0$
since this
is relevant for experiments that have been proposed and for the ongoing SLC
experiment\cite{Ana1}.
We have done the calculations for the $\tau\to\pi\nu_\tau$ and
$\rho\nu_\tau$ channels since these have substantial branching ratios and are
characterized by significant resolving power given by $\alpha_\pi=1$ and
$\alpha_\rho=0.46$. We present the results for $\sqrt{s}=3.67$~GeV, 4.25~GeV,
10.58~GeV and $m_Z$ in Tables~1 and 2. The correlations shown in
Table 1, arising because the initial state is not CP even, may be compared with
the background arising from statistical fluctuations. The values of $\sqrt{
\langle O_2^2 \rangle } $ for the above energies range from about 0.7 to about
10 GeV. Thus, with $10^7$ $\tau$ pairs being produced, the effect of
statistical
background is dominant, and one can ignore the fact that the initial state
is not strictly CP even.

The case with polarization-asymmetrized distributions is somewhat different.
While the values of $\sqrt{\langle O_2^2\rangle }$ continue to be very similar
to the ones in the case of ordinary distributions, the contribution to the
correlation (shown in Table~2) coming
from bremsstrahlung effects is much higher now.  However, the number of
events in the asymmetrized distributions is much less that in the original
distributions. Hence the background arising because the $e^-$ and $e^+$
polarizations are not equal and opposite is again small compared to the
statistical background, and can be neglected.

$\langle O_2\rangle_{\rm SM}$ is small at $\tau$CF energies since parity
violation is not significant and at SLC since the bremsstrahlung process is
non-resonant.

In Ref.~\cite{Cuy} the prospects of probing the dipole form factors with $O_1$
and $O_2$ were evaluated. Here $m_t=175$~GeV and for the planned NLC energy of
$\sqrt{s}=500$~GeV, we have
computed $\langle O_2\rangle_{\rm SM}$ for various
values of $P_e$ ($P_{\bar{e}}=0$) according to
the procedure described.
The results in Table 3 for the bremsstrahlung contributions to the
correlations show that they are negligible compared to
$\sqrt{\langle O_2^2\rangle } /\sqrt{N}$ with
$\sqrt{\langle O_2^2\rangle } \approx 64.5$ GeV as found earlier, and
$\sqrt{N}\approx 600-900$ as expected for an integrated luminosity of
10 fb$^{-1}$.

\section{Conclusions}

We have calculated the $O(\alpha)$ effects due to helicity-flip
bremsstrahlung in
the measurement of CP-violating form factors.
By adapting the equivalent particle approach of Falk and Sehgal to include
polarization, we compute corrections within this picture to
correlations that are enhanced by the introduction of polarization.
We have shown that
besides direct searches for new physics, in indirect ones such as through
measurements of correlations, the bremsstrahlung process plays a potentially
significant role. In practice these remain small as compared to the background
coming from statistical fluctuations, for the luminosities considered. The
effect is small firstly because it arises at
order $\alpha$, and secondly because it depends on parity
violation in the $e^+e^-$ couplings to the gauge boson, which is small
at low $\tau$CF energies where the photon contribution dominates.
At SLC, where the
$Z$ resonance dominates the cross section the effect is even
smaller, because it is a convolution of the Breit-Wigner resonance form with
a bremsstrahlung probability function which is large only away from the peak.
At NLC energies, for the luminosities anticipated, the effect would be
swamped by the statistical fluctuations.

\bigskip

\noindent
{\bf Acknowledgments:}
BA thanks the Swiss National Science Foundation for support
during the course of this work. We thank D.~Rickebusch for assistance with the
manuscript.

\newpage
\section*{Table Captions}

\begin{enumerate}
\item[Table 1(a):] Values of $\langle O_2\rangle_{\rm SM}$ in GeV
for $\pi\pi$, $\pi\rho$
and $\rho\rho$ channels for $\sqrt{s}=3.67$~GeV for chosen values of $P_e$
($P_{\bar{e}}=0$).
\item[Table 1(b):] As above for $\sqrt{s}=4.25$~GeV.
\item[Table 1(c):] As above for $\sqrt{s}=10.58$~GeV.
\item[Table 1(d):] As above for $\sqrt{s}=m_Z$.
\item[Table 2:] Values of $\langle O_2\rangle_{\rm SM}^A$ in GeV
for the polarization
asymmetrized distributions for $\sqrt{s}=3.67$~GeV, 4.25~GeV, 10.58~GeV and
$m_Z$.
\item[Table 3:] Values of $\langle O_2\rangle_{\rm SM}$ in GeV
for the $t\bar{t}$ system
at $\sqrt{s}=500$~GeV for chosen values of $P_e$ ($P_{\bar{e}}=0$).
\end{enumerate}

\newpage
\begin{center}
\[
\begin{array}{||r|r|r|r||}\hline
P_e \quad&\pi\pi\qquad&\pi\rho\qquad & \rho\rho\qquad\\\hline
0\quad &0\qquad &0\qquad &0\qquad \\
-0.62 &-1.0\cdot 10^{-12}&-1.1\cdot 10^{-12}&-1.1\cdot 10^{-12}\\
 0.62 & 1.0\cdot 10^{-12}& 1.1\cdot 10^{-12}& 1.1\cdot 10^{-12}\\
-0.75 &-1.2\cdot 10^{-12}&-1.3\cdot 10^{-12}&-1.4\cdot 10^{-12}\\
 0.75 & 1.2\cdot 10^{-12}& 1.3\cdot 10^{-12}& 1.4\cdot 10^{-12}\\
-1.00 &-1.6\cdot 10^{-12}&-1.7\cdot 10^{-12}&-1.8\cdot 10^{-12}\\
 1.00 & 1.6\cdot 10^{-12}& 1.7\cdot 10^{-12}& 1.8\cdot 10^{-12}\\\hline
\end{array} \]

Table 1(a)

\vspace{1cm}
\[
\begin{array}{||r|r|r|r||}\hline
P_e \quad&\pi\pi\qquad&\pi\rho\qquad & \rho\rho\qquad\\\hline
0\quad &0\qquad &0\qquad &0\qquad \\[2mm]
-0.62 &-6.5\cdot 10^{-10}&-6.5\cdot 10^{-10}&-6.4\cdot 10^{-10}\\
 0.62 & 6.5\cdot 10^{-10}& 6.5\cdot 10^{-10}& 6.4\cdot 10^{-10}\\
-0.75 &-7.9\cdot 10^{-10}&-7.8\cdot 10^{-10}&-7.7\cdot 10^{-10}\\
 0.75 & 7.9\cdot 10^{-10}& 7.8\cdot 10^{-10}& 7.7\cdot 10^{-10}\\
-1.00 &-1.1\cdot 10^{-10}&-1.0\cdot 10^{-10}&-1.0\cdot 10^{-10}\\
 1.00 & 1.1\cdot 10^{-10}& 1.0\cdot 10^{-10}& 1.0\cdot 10^{-10}\\\hline
\end{array} \]

Table 1(b)

\vspace{1cm}
\[
\begin{array}{||r|r|r|r||}\hline
P_e \quad&\pi\pi\qquad&\pi\rho\qquad & \rho\rho\qquad\\\hline
0\quad &0\qquad &0\qquad &0\qquad \\[2mm]
-0.61 &-1.7\cdot 10^{-6}&-1.4\cdot 10^{-6}&-1.1\cdot 10^{-6}\\
 0.61 & 1.7\cdot 10^{-6}& 1.4\cdot 10^{-6}& 1.1\cdot 10^{-6}\\
-0.75 &-2.1\cdot 10^{-6}&-1.7\cdot 10^{-6}&-1.3\cdot 10^{-6}\\
 0.75 & 2.1\cdot 10^{-6}& 1.7\cdot 10^{-6}& 1.3\cdot 10^{-6}\\
-1.00 &-2.8\cdot 10^{-6}&-2.3\cdot 10^{-6}&-1.8\cdot 10^{-6}\\
 1.00 & 2.8\cdot 10^{-6}& 2.3\cdot 10^{-6}& 1.8\cdot 10^{-6}\\\hline
\end{array} \]

Table 1(c)

\vspace{1cm}
\[
\begin{array}{||r|r|r|r||}\hline
P_e \quad&\pi\pi\qquad&\pi\rho\qquad & \rho\rho\qquad\\\hline
0\quad &0\qquad &0\qquad &0\qquad \\[2mm]
-0.62 &-2.3\cdot 10^{-5}&-1.7\cdot 10^{-5}&-1.1\cdot 10^{-5}\\
 0.62 & 2.8\cdot 10^{-5}& 2.1\cdot 10^{-5}& 1.4\cdot 10^{-5}\\
-0.75 &-2.1\cdot 10^{-5}&-2.0\cdot 10^{-5}&-1.4\cdot 10^{-5}\\
 0.75 & 3.4\cdot 10^{-5}& 2.6\cdot 10^{-5}& 1.7\cdot 10^{-5}\\
-1.00 &-3.5\cdot 10^{-5}&-2.6\cdot 10^{-5}&-1.7\cdot 10^{-5}\\
 1.00 & 4.8\cdot 10^{-5}& 3.6\cdot 10^{-5}& 2.4\cdot 10^{-5}\\
\hline
\end{array} \]

Table 1(d)

\vspace{1cm}
\[
\begin{array}{||r|r|r|r||}\hline
\sqrt{s}\quad &\pi\pi\qquad &\pi\rho \qquad& \rho\rho\qquad\\[2mm]\hline
 3.67 & 1.8\cdot 10^{-8}& 1.9\cdot 10^{-8}& 2.0\cdot 10^{-8}\\
 4.25 & 8.8\cdot 10^{-6}& 8.7\cdot 10^{-6}& 8.6\cdot 10^{-6}\\
 10.58 & 3.7\cdot 10^{-3}& 3.0\cdot 10^{-3}& 2.3\cdot 10^{-3}\\
 m_Z\; &-2.6\cdot 10^{-4}&-1.9\cdot 10^{-4}&-1.3\cdot 10^{-4}\\\hline
\end{array} \]

Table 2

\vspace{1cm}
\[
\begin{array}{||r|r||}\hline
P_e \quad& \langle O_2\rangle_{{\rm SM}}\quad \\[2mm]\hline
-0.62 & 1.5\cdot 10^{-3}\\
 0.62 &-2.4\cdot 10^{-3}\\
-0.75 & 1.7\cdot 10^{-3}\\
 0.75 &-2.9\cdot 10^{-3}\\
-1.00 & 2.2\cdot 10^{-3}\\
 1.00 &-4.8\cdot 10^{-3}\\\hline
\end{array} \]

Table 3
\end{center}
\end{document}